\documentclass[prc,twocolumn,showpacs]{revtex4}
\usepackage{graphicx}
\usepackage{dcolumn}
\usepackage{bm}

\def\bvec#1{\mbox{\boldmath $#1$}}

\newcommand{\larw}[1]{\overleftarrow{#1}}
\newcommand{\rarw}[1]{\overrightarrow{#1}}

\newcommand{\bra}{\langle}
\newcommand{\ket}{\rangle}

\begin{document}

\title{Subtraction of the spurious translational mode 
from the RPA response function}

\author{Kazuhito Mizuyama$^{1}$}
\email{mizukazu147@gmail.com}
\author{Gianluca Col\`o$^{2,1}$}
\email{colo@mi.infn.it}

\affiliation{
$^{1}$ 
INFN, Sezione di Milano, 
via Celoria 16, 20133 Milano, Italy
\\
$^{2}$ 
Dipartimento di Fisica, Universit$\grave{a}$ degli 
Studi di Milano, via Celoria 16, 20133 Milano, Italy
}


\date{\today}

\begin{abstract}
It is well known that within self-consistent Random
Phase Approximation (RPA) on top of Hartree-Fock (HF),
the translational symmetry should be restored. Due
to approximations at the level of the practical 
implementation, this restoration may be only partial. 
As a result, one has spurious contributions
in the physical quantities that are extracted from RPA.
While there are several recipes in the literature
to overcome this drawback in order to produce transition densites
or strength functions that are free from spurious
contamination, there is no formalism associated
with the full RPA response function. We present such
formalism in this paper. Our goal is to avoid spurious
contamination when the response function is used
in many-body frameworks like the particle-vibration
coupling theory. 
\end{abstract}

\maketitle

\section{Introduction}

RPA is one of the basic tools to study the nuclear
vibrational excitations. The basic theory is well
known, and several versions have been
implemented since the early days of nuclear structure
studies. However, while in the past 
the underlying mean field and the residual particle-hole 
(p-h) interactions were often purely empirical, nowadays
self-consistent HF plus RPA calculations are
feasible with quite sophisticated effective forces. 
The importance of these calculations stems from the
possibility of testing their predictive power, e.g.,   
when going towards neutron-rich (or neutron-deficient)
isotopes that can be studied at present or
future radioactive beam facilities. 

One of the advantages of RPA is that it is a
conserving theory, that is, the symmetries of the 
Hamiltonian that are broken at the HF level  
are in principle restored in RPA. The detailed
way in which this restoration is realized has been
already widely discussed in the literature. The
reader can consult standard textbooks (cf. Refs.  
\cite{Ring-Schuck,Blaizot-Ripka}) or review papers like 
Ref. \cite{Lane}. The purpose of the present
work is not to re-discuss basic questions. We
shall focus instead on practical issues, in
the case of the restoration of 
translational symmetry (which is associated
with the appearance of a zero-energy translational 
mode in the RPA $J^\pi=1^-$ spectrum).  

In practice, it is impossible to obtain this
translational mode at {\em exactly} zero energy
due to some theoretical or numerical approximation. 
If we restrict ourselves to the case of nonrelativistic
calculations performed with effective Skyrme
forces, no fully self-consistent calculations
that treat the continuum exactly are available.
Fully self-consistent calculations in which the
continuum is discretized do exist \cite{Terasaki1,Colo2},
and in this case the energy of the spurious 
translational mode goes to zero as the dimension of
the box in which the system is set increases. 
Both in this case, and in the case of continuum
calculations which are the object of our study, 
one has then to deal with the spurious contamination 
of several quantities.

In fact, if the spurious state from a numerical
calculation does not lie at zero energy,
and its wavefunction does not coincide with the
true translational wavefunction, the other (physical)
states contain some admixture of the ideal spurious
state. These admixtures should be eliminated. Various
procedures have been proposed, depending on the
specific RPA implementation. In the case of
coordinate-space RPA the basic question is how
to remove spurious admixtures from the strength
function associated with a given operator 
$\hat F({\bf r})$ acting in the 1$^-$ channel, 
namely from 
\begin{equation}
S(E) = \sum_n \vert \langle n \vert \hat F
\vert \tilde 0 \rangle \vert^2 \delta(E-E_n),
\end{equation}
where we suppose that $\vert n\rangle$ are
the RPA states having respectively energy $E_n$ and
$\vert \tilde 0\rangle$ is the nuclear ground-state.
We assume that the definition can be formally
extended to the continuum case, when the spectrum is
not discrete. Sometimes one is willing to
remove the spurious admixtures directly 
from the transition densities, that are defined as
\begin{equation}
\delta\rho_n({\bf r}) = 
\langle n \vert \hat\rho({\bf r}) \vert \tilde
0 \rangle, 
\label{td}
\end{equation}
where $\hat\rho$ is the density operator. Once more,
this definition holds if the spectrum is continuum
and the transition density is associated with
a state at every positive energy $\omega$: we
will use the notation $\delta\rho(\bvec{r}\omega)$
in what follows.

These two problems have been dealt with extensively in
the literature (see Refs. \cite{Giai1,Suzuki,Sagawa,Colo1,
Hamamoto,Shlomo1,Terasaki1}). The solutions include
projection methods, use of corrected effective operators
etc. However, the present work is motivated by
the use of the dipole RPA spectrum not by itself,
but as an input to particle-vibration coupling (PVC) 
calculations \cite{BM,Mahaux,Colo10}. In PVC calculations
performed in coordinate space with proper continuum
coupling \cite{Mizuyama_tbp} the full RPA response
function enters; consequently, one must deal with
the subtraction of spurious contamination from that
quantity. Since this has not been discussed in
the literature previously, to our best knowledge, 
we discuss the problem in the present paper. The
next Section deals with the theoretical formalism. 

\section{Method}

As recalled in the Introduction, the spurious 
dipole mode is originated by the restoration 
of the translational symmetry in RPA. Translations
along one direction, say the $z$-axis, 
are generated by the total momentum operator
$\hat P$ which is a conserved quantity, i.e., $[H,\hat P]=0$.
As discussed in \cite{Ring-Schuck,Blaizot-Ripka,Thouless}, 
the ideal spurious state must have $X$ and $Y$ 
amplitudes associated with the p-h matrix
elements of $\hat P$. Because of the symmetry of the
RPA matrix, its non-zero eigenvalues and eigenvectors 
appear in pairs. The partner of the spurious state
has amplitudes related to the matrix element of
the operator $\hat Q$ that satisfies $[H,\hat Q]
=\frac{-i\hbar}{Am}\hat P$. This latter equation
is associated with the Galilean invariance as
it is shown in Refs. \cite{Ring-Schuck,Blaizot-Ripka}.

In this Section, we show a method which removes the 
spurious contribution from the RPA response function 
in coordinate space representation. We express the formulas
that have been introduced in Ref. \cite{Nakatsukasa} 
in coordinate 
space representation, and extend them to case 
of the RPA response 
function.

In order to remove spurious admixtures, first of all, 
as in Ref. \cite{Nakatsukasa}, we suppose 
the validity of 
\begin{eqnarray}
\delta\hat{\rho}(\omega)
=
\delta\hat{\tilde{\rho}}(\omega)
+
\lambda_P(\omega)\delta\hat{\rho}_P
+
\lambda_Q(\omega)\delta\hat{\rho}_Q,
\label{asum1}
\end{eqnarray}
together with the orthogonality conditions 
$
[\delta\hat{\tilde{\rho}}(\omega),\delta\hat{\rho}_P]
=[\delta\hat{\tilde{\rho}}(\omega),\delta\hat{\rho}_Q]=0
$, for the ``physical'' transition density
operator $\delta\hat{\tilde{\rho}}(\omega)$
in its second-quantized form. 
$\delta\hat\rho(\omega)$ is the second-quantized form
of the RPA transition density operator. 
$\delta\hat{\rho}_P$ and $\delta\hat{\rho}_Q$ 
are related with $\hat{P}$ and $\hat{Q}$ by
\begin{eqnarray}
\delta\hat{\rho}_P
&=&
i
\sum_{q,\sigma}
\int d\bvec{r}
\hat{\psi}_q^\dagger(\bvec{r}\sigma)(-i\hbar\nabla_z)\hat{\psi}_q(\bvec{r}\sigma)
=i\hat{P}, 
\\
\delta\hat{\rho}_Q
&=&
i
\sum_{q,\sigma}
\int d\bvec{r}
\hat{\psi}_q^\dagger(\bvec{r}\sigma)z\hat{\psi}_q(\bvec{r}\sigma)
=i\hat{Q}.
\end{eqnarray}
The canonicity conditions 
$[\delta\hat{\rho}_Q,\delta\hat{\rho}_P]=-[\hat Q,\hat P]=-i\hbar A$
hold. 

The transition density $\delta\rho(\bvec{r}\omega)$ in coordinate space representation obeys
\begin{eqnarray}
\delta\rho_{q,\alpha}(\bvec{r}\omega)
=
\bra 0|[\hat{\rho}_{q,\alpha}(\bvec{r}),\delta\hat{\rho}(\omega)]|0\ket, 
\label{exp1}
\end{eqnarray}
where $\hat{\rho}_{q,\alpha}(\bvec{r})$ 
is the density operator defined by
\begin{eqnarray}
\hat{\rho}_{q,\alpha}(\bvec{r})
&=&
\sum_\sigma
\hat{\psi}^\dagger_q(\bvec{r}\sigma)
\hat{O}_\alpha(\bvec{r})
\hat{\psi}_q(\bvec{r}\sigma)
\nonumber\\&&
\mbox{with }
\hat{O}_\alpha(\bvec{r})\in \{1,\rarw{\nabla}\pm\larw{\nabla},...\}.
\label{densop}
\end{eqnarray}
Using Eq. (\ref{asum1}) and Eq. (\ref{exp1}), 
we obtain
\begin{eqnarray}
\delta\rho_{q,\alpha}(\bvec{r}\omega)
&=&
\delta\tilde{\rho}_{q,\alpha}(\bvec{r}\omega)
+
\lambda_P(\omega)
\bra 0|[\hat{\rho}_{q,\alpha}(\bvec{r}),\delta\hat{\rho}_P]|0\ket
\nonumber\\&&
+
\lambda_Q(\omega)
\bra 0|[\hat{\rho}_{q,\alpha}(\bvec{r}),\delta\hat{\rho}_Q]|0\ket.
\label{exp2}
\end{eqnarray}
By using Eq. (\ref{asum1}) and Eq. (\ref{exp1}) with the orthogonality condition 
and the canonicity condition, $\lambda_P(\omega)$ and $\lambda_Q(\omega)$ are 
represented by the transition densities in coordinate space representation as 
\begin{eqnarray}
\lambda_P(\omega)
&=&
\frac{\bra 0|[\delta\hat{\rho}_Q,\delta\hat{\rho}(\omega)]|0\ket}
{\bra 0|[\delta\hat{\rho}_Q,\delta\hat{\rho}_P]|0\ket}
=
\frac{-1}{\hbar A}
\sum_q\int d\bvec{r} z \delta\rho_q(\bvec{r}\omega),
\nonumber\\
\label{coef1}
\\
\lambda_Q(\omega)
&=&
-\frac{\bra 0|[\delta\hat{\rho}_P,\delta\hat{\rho}(\omega)]|0\ket}
{\bra 0|[\delta\hat{\rho}_Q,\delta\hat{\rho}_P]|0\ket}
=
\frac{1}{A}
\sum_q\int d\bvec{r} \delta j_{z,q}(\bvec{r}\omega).
\nonumber\\
\label{coef2}
\end{eqnarray}
For example, if $\hat{O}_\alpha(\bvec{r})=1$, then Eq. (\ref{exp2}) becomes 
\begin{eqnarray}
\delta\rho_q(\bvec{r}\omega)
&=&
\delta\tilde{\rho}_q(\bvec{r}\omega)
\nonumber\\
&&
+
\nabla_z\rho_q(\bvec{r})
\frac{-1}{A}
\sum_{q'}\int d\bvec{r}' z' \delta\rho_{q'}(\bvec{r}'\omega)
\label{12}
\end{eqnarray}
because $\bra 0|[\hat{\rho}_q(\bvec{r}),\delta\hat{\rho}_P]|0\ket
=\hbar\nabla_z\rho_q(\bvec{r})$.
As a further example, if $\hat{O}_\alpha(\bvec{r})
=\rarw{\nabla}-\larw{\nabla}$, one has
\begin{eqnarray}
\delta 2i\bvec{j}_q(\bvec{r}\omega)
&=&
\delta 2i\tilde{\bvec{j}}_q(\bvec{r}\omega)
\nonumber\\
&&
+
\rho_q(\bvec{r})
\frac{\hat{e}_z}{2A}
\sum_{q'}\int d\bvec{r}' \delta 2i j_{z,q'}(\bvec{r}'\omega)
\nonumber \\
\end{eqnarray}
because $\bra 0|[2i\hat{\bvec{j}}_q(\bvec{r}),\delta\hat{\rho}_Q]|0\ket
=i\hat{e}_z\rho_q(\bvec{r})$ ($\hat{e}_z$ is the unit vector
along $z$).

In this paper, 
we extend the above method to single out ``physical'' 
quantities labelled by a tilde, to the case of the response function instead
of the transition density.
We adopt here the same notation as in Ref. 
\cite{mizuyama}. The response function $R^{\alpha\beta}_{qq'}$
depends on two indices $\alpha$ and $\beta$ that identify the
operators as in Eq. (\ref{densop}). 

As in Ref. \cite{mizuyama}, one can write
\begin{eqnarray}
\delta\rho_q(\bvec{r}\omega)&=&\sum_{q'}
\int d\bvec{r}'R^{1,1}_{qq'}(\bvec{r}\bvec{r}';\omega)f_{q'}(\bvec{r}'),
\label{resp1} 
\\
\delta 2i\bvec{j}_q(\bvec{r}\omega)&=&\sum_{q'}
\int d\bvec{r}'R^{\nabla_-,1}_{qq'}(\bvec{r}\bvec{r}';\omega)f_{q'}(\bvec{r}'),
\label{resp2}
\end{eqnarray}
where $f(\bvec{r})$ is the external operator. In the
previous formula the superscript 1 refers to the identity
operator and the superscript $\nabla_-$ to the 
operator $\overrightarrow{\nabla}-\overleftarrow{\nabla}$. 

We can use Eqs. (\ref{12}) and (\ref{resp1}); then,  
for the response function we can arrive at
\begin{eqnarray}
&&\tilde{R}^{11}_{qq'}(\bvec{r}\bvec{r}';\omega)
=
R^{11}_{qq'}(\bvec{r}\bvec{r}';\omega)
\nonumber\\
&&
-
\nabla_z\rho_q(\bvec{r})
\frac{-1}{A}
\sum_{q''}\int d\bvec{r}'' z'' 
R^{11}_{q''q'}(\bvec{r}''\bvec{r}';\omega).
\label{res1}
\end{eqnarray}
This is our main result. It allows extracting the
``physical'' response function $\tilde{R}$ from the RPA result
$R$.

In the next Section, we illustrate our method by means of 
some numerical results for the strength function $S$. In fact, 
by using the ``physical'' response function defined in 
Eq. (\ref{res1}), the ``physical'' strength function 
can be written as
\begin{eqnarray}
\tilde{S}(\omega)
&=&
-\frac{1}{\pi}\mbox{Im}
\sum_{qq'}
\int\!\!\int\!\! d\bvec{r}d\bvec{r}'
f_q^*(\bvec{r})
\tilde{R}^{11}_{qq'}(\bvec{r}\bvec{r}';\omega)
f_{q'}(\bvec{r}')
\nonumber\\
&=&
S(\omega)\nonumber\\
&&
-
\frac{1}{\pi}
\mbox{Im}
\sum_{q}
\int\!\!d\bvec{r}
f_q^*(\bvec{r})
\nabla_z\rho_q(\bvec{r})/A
\nonumber\\
&&\times
\left[
\sum_{q'q''}
\int\!\!d\bvec{r}'
\int d\bvec{r}'' 
z'' 
R^{11}_{q''q'}(\bvec{r}''\bvec{r}';\omega)
f_{q'}(\bvec{r}')
\right]
\nonumber\\
&=&
S(\omega)
-
\delta S(\omega), 
\label{strength}
\end{eqnarray}
where $S(\omega)$ is the strength function defined by 
the RPA without corrections and $\delta S(\omega)$ 
is the spurious contribution. 

As mentioned in the Introduction, the main point of
correcting the response function instead of the
transition densities or transition strength is that
the response function is an input for more sophisticated
many-body frameworks like the particle-vibration coupling
approach. In this case, 
the static HF potential is generalized to a time-dependent
(or energy-dependent) self-energy function.
This self-energy function $\Sigma$ is associated
with the coupling of single-nucleon states with RPA
vibrations; consequently, it can be defined in terms of 
the HF Green's function $G_0$ and the RPA response function. 
This definition can be found in the references mentioned
in the Introduction. However, we have recently tried
for the first time \cite{Mizuyama_tbp} to formulate
this theory in coordinate space. In this case, 
in order to remove the contribution of the spurious state by using 
Eq. (\ref{res1}), the self-energy function must be expressed as
\begin{eqnarray}
&&
\tilde{\Sigma}(\bvec{r}\sigma,\bvec{r}'\sigma';\omega)
\nonumber\\
&&=
\int_{-\infty}^{\infty}\frac{d\omega'}{2\pi}
\kappa(\bvec{r})
G_0(\bvec{r}\sigma,\bvec{r}'\sigma';\omega-\omega')
\kappa(\bvec{r}')
i\tilde{R}^{11}(\bvec{r}\bvec{r}';\omega'),
\nonumber\\
\label{selfen}
\end{eqnarray}
where $\kappa$ is the residual (p-h) interaction.

Using this self-energy function, we can solve the Dyson equation 
$G=G_0+G_0\tilde{\Sigma}G$ in coordinate space, and then obtain 
the perturbed Green's function. The detail of the formulas will 
be in our forthcoming paper \cite{Mizuyama_tbp}, where also
several results are discussed. In the present work, we
present some few results about the relevance of
projecting out the spurious contribution.

\section{Numerical results}

In this Section, we study the strength functions both in the case 
of the electric dipole and the isoscalar compression operators. 
For this study, the nucleus of choice is $^{208}$Pb. 
The equations are solved in 
a radial mesh that extends up to 20 fm. The 
radial step is 0.2 fm. The states up to angular momentum
$\ell$=12 are included in the model space. The Skyrme
set employed is SLy5 \cite{Chabanat}. The implementation
of the continuum RPA is exactly as in Ref. \cite{mizuyama}. 

Also we apply our method to Hartree-Fock plus particle-vibration 
coupling. As an example, we show the level density in the case of 
$^{24}$O. 
We included the RPA electric dipole phonons up 
to 60 MeV. We discuss the spurious state contribution to the level 
density, extracted from the Green's function which
is solution of the Dyson equation (in the level density we adopt 
a smoothing 
parameter equal to 0.4 MeV).

\subsection{E1 excitations}

\begin{figure}[htbp]
\includegraphics[width=\textwidth,angle=-90,scale=0.5]{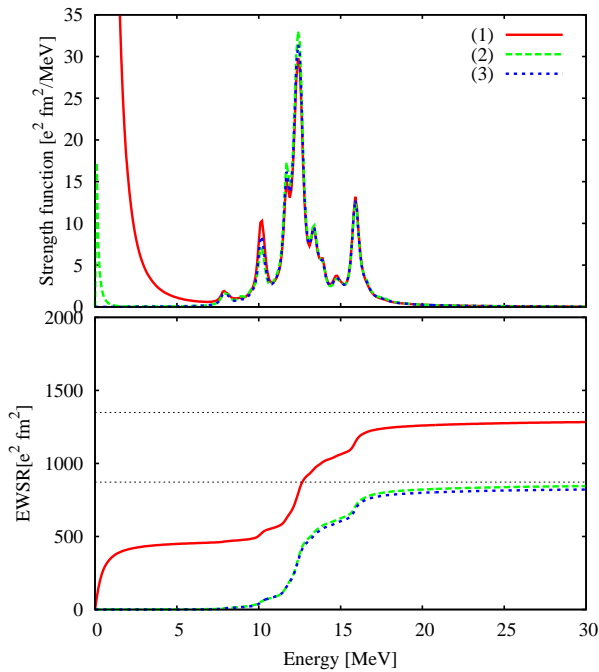}
\caption{(Color online) The Skyrme continuum RPA results for 
the E1 strength of $^{208}$Pb 
with the SLy5 Skyrme interaction are displayed in the upper panel. 
The red curve [labelled by (1) in the 
figure] is calculated by using the E1 operator (\ref{e1op1}), whereas 
the green curve (2) is calculated by using the E1 operator with
the center-of-mass subtracted [Eq. (\ref{e1op2})], and 
the blue curve (3) is calculated by using the operator of Eq. (\ref{e1op1}) 
but at the same time with the spurious state
eliminated from the response function by using our method. 
In the lower pannel we show the running (or cumulated) 
values of the energy weighted sum rules. The dotted lines 
are the double-commutator values of the energy-weighted sum rules 
for the operators (\ref{e1op1}) and (\ref{e1op2}). The formulas
are given in the Appendix [Eqs. (\ref{A4}) and (\ref{e1ewsr1}), 
respectively] and the values are shown in 
Table \ref{ewsrtable}.}
\label{fige1}
\end{figure}

In Fig.~\ref{fige1}, we display the strength functions 
(and the associated energy-weighted sum rules, EWSRs) 
obtained by using the following E1 operators:
\begin{eqnarray}
F_{E1}
=e\sum_{p=1}^Zr_p Y_{1M}(\hat{\bvec{r}}_p),
\label{e1op1}
\end{eqnarray}
and
\begin{eqnarray}
\tilde{F}_{E1}
=e\sum_{n=1}^N \frac{Z}{A}r_n Y_{1M}(\hat{\bvec{r}}_n)
-e\sum_{p=1}^Z \frac{N}{A}r_p Y_{1M}(\hat{\bvec{r}}_p).
\label{e1op2}
\end{eqnarray}
The operator defined by Eq. (\ref{e1op2}) is obtained by
removing the center of mass 
from the original operator of Eq. (\ref{e1op1}). 
We apply our method to obtain $\tilde S(\omega)$, by
leaving the operator equal to 
(\ref{e1op1}), and we obtain the blue curve in
Fig. \ref{fige1}. One can compare this result to the E1 spectrum 
without the spurious  
subtraction (red curve), and to the one obtained by using 
the original RPA response and the operator (\ref{e1op2}) 
(green curve). 
In the lower panel of the figure we show the running, or
cumulated, values of EWSRs.
From both panels one can argue that  
our method succeeds to remove the spurious state. In
particular, the success of our microscopic implementation 
is very clear from the detailed strength of the upper panel, as
no strength remains in the low-energy region down to zero.  

\subsection{IS compression dipole}

Similar accurate results can be obtained if the 
IS dipole compression operator is applied. This operator reads
\begin{eqnarray}
F_{IS1}
=\sum_{i=1}^A r^3_i Y_{1M}(\hat{\bvec{r}}_i)
\label{cL1op1}
\end{eqnarray}
without any center-or-mass subtraction. Many authors
have used, to take care of this subtraction, the 
hydrodynamical ansatz
that has been originally proposed in Ref. \cite{Giai1}.
This amounts to using the operator
\begin{eqnarray}
\tilde{F}_{IS1}
=\sum_{i=1}^A \left(r^3_i-\frac{5}{3}\bra r^2\ket r_i\right) 
Y_{1M}(\hat{\bvec{r}}_i),
\label{cL1op2}
\end{eqnarray}
the factor $5\langle r^2 \rangle/3$ being often denoted 
with the symbol $\eta$. 

\begin{figure}[htbp]
\includegraphics[width=\textwidth,angle=-90,scale=0.5]{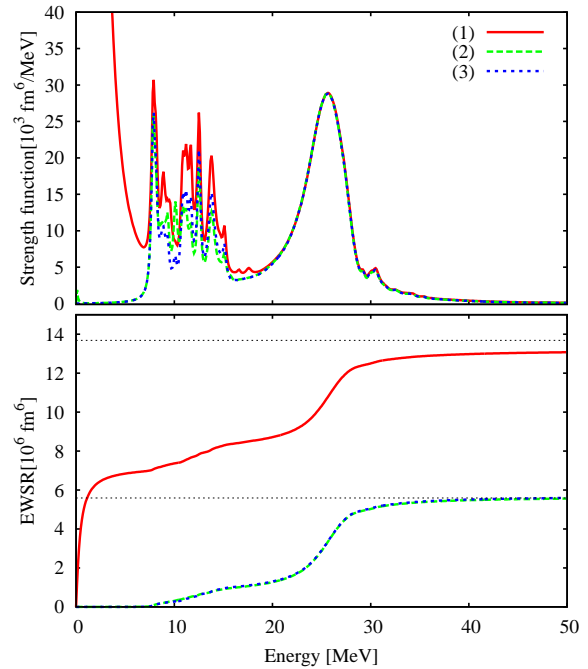}
\caption{(Color online) Same as Fig.\ref{fige1} 
for the isoscalar compression dipole spectrum. 
The nucleus is $^{208}$Pb and the Skyrme set is SLy5. 
The red curve [labelled by (1) in the
figure] is calculated by using the operator (\ref{cL1op1}), whereas
the green curve (2) is calculated by using the operator with
the center-of-mass subtracted [Eq. (\ref{cL1op2})], and
the blue curve (3) is calculated by using the operator Eq. (\ref{cL1op1})
but at the same time with the spurious state
eliminated from the response function by using our method.
In the lower pannel we show the running (or cumulated)
values of the energy weighted sum rules. The dotted lines
are the double-commutator values of the energy-weighted sum rules
for the operators (\ref{cL1op1}) and (\ref{cL1op2}). The formulas
are given in the Appendix and the values are shown in
Table \ref{ewsrtable}.}
\label{figcL1}
\end{figure}

We present our results in a form which is analogous to
the one of the previous subsection in Fig. \ref{figcL1}. 
The blue line corresponds to the choice of the bare
operator (\ref{cL1op1}), and to the miscroscopic
subtraction of the center-of-mass motion proposed
in our present work. The associated result can be
compared with the use of the operator (\ref{cL1op2}) 
in connection with the original RPA response 
(green curve) and to the RPA result without
any subtraction of the center-of-mass either in the
operator or in the response function (red curve). 

The result of the red curve includes the spurious
contributions. Both other curves provide a way to
remove these contributions. Although similar, they
are not identical: in this respect, the upper panel
of Fig. \ref{figcL1} shows to which extent the
ansatz (\ref{cL1op2}) for the effective operator
is accurate. The small differences are not visible
if one looks only at the cumulated EWSRs: both the
blue and green line converge to the same result.

\begin{table}
\renewcommand\arraystretch{1.5}
\begin{tabular}{ccccc}
\hline
\hline
& Operator & $\tilde{m_1}$ & $m_1$  & $\delta m_1$ \\
\hline
E1                & (\ref{e1op1})     & 872.3 & 1350  & 477.7\\
                  & (\ref{e1op2})     & 872.3 & 872.3 & 0.0\\
Compression dip.  & (\ref{cL1op1}) & $5.588\times 10^6$ & $1.368\times 10^7$ & $8.092\times 10^6$\\
                  & (\ref{cL1op2}) & $5.588\times 10^6$ & $5.588\times 10^6$ & 0.0\\
\hline
\end{tabular}
\caption{Values of the energy-weighted sum rule for the E1 and the isoscalar compression
dipole excitations in $^{208}$Pb with SLy5. These values correspond to the dotted lines in Fig. \ref{fige1}
and \ref{figcL1}. $\delta m_1$ is the contribution arising from the center-of-mass 
spurious motion. Definition are given in the Appendix: in particular $\tilde{m}_1$ is defined by Eq. (\ref{m1}).}
\label{ewsrtable}
\end{table}

\subsection{Spurious mode subtraction in PVC}

The level density $\rho_{lj}$ can be defined as
\begin{eqnarray}
\rho_{lj}(\omega)
=
\frac{\pm 1}{\pi}\int_0^\infty dr 
\mbox{Im} 
G_{lj}(rr,\omega),
\label{lvd}
\end{eqnarray}
where
$G_{lj}$ can be either the Hartree-Fock Green's function or the perturbed Green's function 
which is obtained by solving the Dyson equation. The spurious state may affect 
the self-energy function that appears in the Dyson 
equation because the definition includes 
the RPA response function. 
Therefore, the spurious contribution should be systematically
eliminated from the response function.

\begin{figure}[htbp]
\includegraphics[width=\textwidth,angle=-90,scale=0.28]{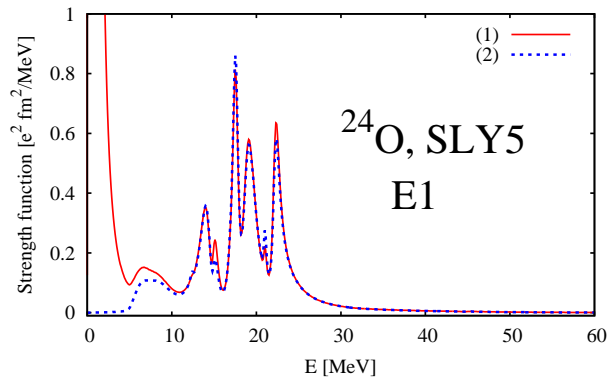}
\caption{(Color online) Same as Fig. \ref{fige1} for $^{24}$O. 
The solid red curve [labelled by (1) in the figure] is calculated by using the 
operator (\ref{e1op1}), and the dotted blue curve (2) is calculated by using the 
operator Eq. (\ref{e1op1}) but at the same time with the spurious state eliminated 
from the response function by using our method.}
\label{fige1O24}
\end{figure}

In the neutron-rich nuclei like $^{24}$O, low-lying 
E1 strength can be larger than in other systems, and the spurious state 
is not well separated from 
the other physical states, as it is shown in Fig. \ref{fige1O24}.

\begin{figure}[htbp]
\includegraphics[width=\textwidth,angle=-90,scale=0.4]{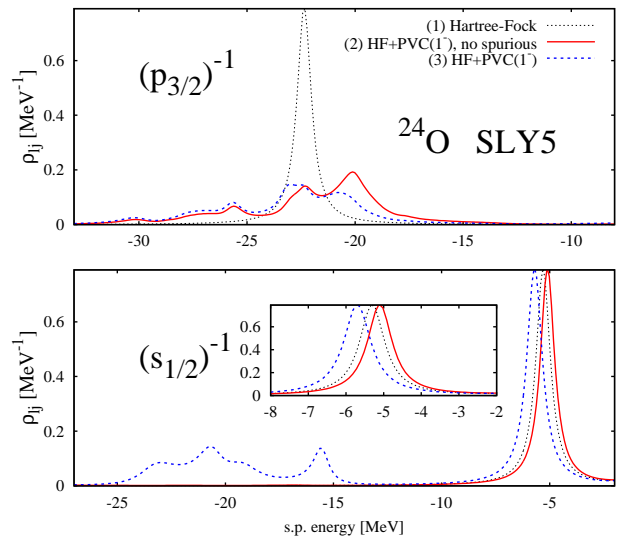}
\caption{(Color online) The level density defined by Eq. (\ref{lvd}) 
in $^{24}$O. 
The dotted black curve [labelled by (1) in the figure] is the Hartree-Fock level 
density. The solid red curve [labelled by (2) in the figure] is the HF+PVC level 
density when the spurious state is removed by applying Eq. (\ref{res1}). 
The dashed blue curve [labelled by (3) in the figure] is also 
the HF+PVC level density, but when the spurious state is not removed. 
The upper panel corresponds to  
$(p_{3/2})^{-1}$, and the lower panel is $(s_{1/2})^{-1}$ orbit.}
\label{lvdO24}
\end{figure}

In Fig.\ref{lvdO24}, the level density is plotted as a function of 
the single particle energy. The dotted black 
curve is the Hartree-Fock level density. 
The solid red curve is the HF+PVC level density, 
in which the spurious contribution is properly removed. 
The dashed blue curve is the HF+PVC level density, but when 
the spurious state is not removed. We have studied neutron level
densities, and in particular 
the upper pannel is the level density in the $(p_{3/2})^{-1}$ case, 
and the lower pannel is 
the $(s_{1/2})^{-1}$ case. 
By comparing the solid red curve and the dashed 
blue curve, we can see that the unphysical contribution of 
the spurious state cannot be ignored.

\section{Summary}

While RPA provides, in principle, a theoretical framework that restores
the symmetries which are broken at the mean field level, approximations that
are often made can spoil this feature. This is particularly true for the
case of the translational symmetry. It has been known for several decades
that the self-consistency violations, the cutoff in the model space and/or
other limitations of the practical RPA implementations leave the spurious
center-of-mass motion at finite instead of zero energy and give
associated contaminations of the wavefunctions and physical observables
related with the RPA states.

Many papers have been devoted to this problem, yet the subject of this
work is the removal of spurious contributions from the RPA full
response function, within the continuum RPA on top of Skyrme HF framework.
We have presented a microscopic formalism that is claimed to be able to achieve this
removal quite successfully, and we have validated this statement by means
of accurate calculations of the E1 and isoscalar compression spectra
of $^{208}$Pb.

The original motivation of this work is the possibility of using the
RPA response function for particle-vibration coupling calculations.
Indeed we have shown, by using the example of level densities 
in $^{24}$O, that the results of PVC either with or without 
the proper subtraction of the spurious state can substantially
differ. 
Other many-body scheme or reaction calculations require the
use of the response function. In all these cases, our proposed method
can display its usefulness.

\appendix
\section{Energy-weighted sum rules of interest}

In general, we define the energy-weighted sum rule associated 
with the strength function $S(\omega)$ as
\begin{equation}
m_1=\int d\omega \omega S(\omega).
\end{equation}
Consistently with Eq. (\ref{strength}) we define
\begin{equation}
\tilde{m}_1=
\int d\omega \omega\tilde{S}(\omega)=
m_1 - \delta m_1
\label{m1}
\end{equation}
with 
\begin{equation}
\delta m_1=\int d\omega \omega\delta S(\omega).
\end{equation}
It is well known that $m_1$ can be calculated analitically by means
of the expectation value of the double commutator. In this Appendix, we
only summarize the useful results in the case of Skyrme interactions. 
 
\subsection{The E1 operator}

The $m_1$ associated with the operator (\ref{e1op1}) is given by
\begin{eqnarray}
m_1=
\frac{e^2\hbar^2}{2m}\frac{9}{4\pi}Z
+
\frac{9}{4\pi}e^2b I,
\label{A4}
\end{eqnarray}
with
\begin{eqnarray}
b&=&\frac{1}{4}\left(t_1(1+\frac{1}{2}x_1)+t_2(1+\frac{1}{2}x_2)\right),
\\
I&=&\int dr 4\pi r^2 \rho_n(r)\rho_p(r).
\end{eqnarray}
The second term of $m_1$ is a correction arising from the momentum-dependent 
terms of the Skyrme interaction.

$\delta m_1$ is given by
\begin{eqnarray}
\delta m_1=\frac{e^2\hbar^2}{2m}\frac{9}{4\pi}\frac{Z^2}{A}.
\end{eqnarray}
Therefore, 
\begin{eqnarray}
\tilde{m}_1
&=&
m_1 - \delta m_1
\nonumber\\
&=&
\frac{e^2\hbar^2}{2m}\frac{9}{4\pi}Z
+
\frac{9}{4\pi}e^2b I
-
\frac{e^2\hbar^2}{2m}\frac{9}{4\pi}\frac{Z^2}{A}
\nonumber\\
&=&
\frac{e^2\hbar^2}{2m}\frac{9}{4\pi}\frac{NZ}{A}
+
\frac{9}{4\pi}e^2b I.
\label{e1ewsr1}
\end{eqnarray}

The $m_1$ associated with the operator (\ref{e1op2}) is given by
\cite{terasaki,Shlomo2}
\begin{eqnarray}
m_1=
\frac{e^2\hbar^2}{2m}\frac{9}{4\pi}\frac{NZ}{A}
+
\frac{9}{4\pi}e^2b I,
\label{e1ewsr2}
\end{eqnarray}
and, as expected, Eq. (\ref{e1ewsr1}) and (\ref{e1ewsr2}) are 
completely equivalent. 

\subsection{The compression dipole operator}

The $m_1$ with the operator (\ref{cL1op1}) is given by 
\begin{equation}
m_1=
\frac{\hbar^2}{2m}\frac{3}{4\pi}A
\bra 11r^4 \ket.
\end{equation}
Also $\delta m_1$ can be derived easily as
\begin{eqnarray}
\delta m_1
=\frac{\hbar^2}{2m}A \frac{1}{4\pi}\bra 5r^2 \ket\bra 5r^2 \ket, 
\end{eqnarray}
and therefore 
\begin{eqnarray}
\tilde{m}_1
&=&
m_1 - \delta m_1
\nonumber\\
&=&
\frac{\hbar^2}{2m}\frac{3}{4\pi}A
\bra 11r^4 \ket
-
\frac{\hbar^2}{2m}\frac{1}{4\pi}A
\bra 5r^2 \ket^2
\nonumber\\
&=&
\frac{\hbar^2}{2m}\frac{1}{4\pi}A
\left(
33\bra r^4 \ket-
25\bra r^2 \ket^2
\right).
\label{A12}
\end{eqnarray}
On the other hand, the energy-weighted sum rule $m_1$ associated with (\ref{cL1op2})
reads
\begin{eqnarray}
m_1=\frac{\hbar^2}{2m}\frac{1}{4\pi}A
\left(
33\bra r^4 \ket-
25\bra r^2 \ket^2
\right).
\label{A13}
\end{eqnarray}
Once more, Eq. (\ref{A12}) and (\ref{A13}) are 
completely equivalent. 


\end{document}